\documentclass[conference]{IEEEtran}
\IEEEoverridecommandlockouts
\usepackage{cite}
\usepackage{amsmath,amssymb,amsfonts}
\usepackage{algorithmic}
\usepackage{graphicx}
\usepackage{textcomp}
\usepackage{xcolor}
\usepackage{booktabs}
\usepackage{multirow}
\usepackage{array}
\def\BibTeX{{\rm B\kern-.05em{\sc i\kern-.025em b}\kern-.08em
    T\kern-.1667em\lower.7ex\hbox{E}\kern-.125emX}}

\graphicspath{{figures/}}

\begin{document}

\title{An Edge-Cloud Collaborative Architecture for Proactive Elderly Care:\\Real-Time Risk Assessment and Three-Level Emergency Response}

\author{\IEEEauthorblockN{Lijie ZHOU}
\IEEEauthorblockA{\textit{School of Computer Science}\\
\textit{University of Nottingham Ningbo China}\\
Ningbo, China\\
scylz12@nottingham.edu.cn}
\and
\IEEEauthorblockN{Luran WANG}
\IEEEauthorblockA{\textit{School of Mathematical Sciences}\\
\textit{University of Nottingham Ningbo China}\\
Ningbo, China\\
smylw4@nottingham.edu.cn}}

\maketitle

\begin{abstract}
The rapid aging of global populations has created an urgent need for intelligent healthcare monitoring systems that can ensure the safety and well-being of elderly individuals living independently. Existing cloud-centric elderly care platforms suffer from significant limitations, including high latency that is unacceptable for emergency response, privacy concerns due to continuous transmission of sensitive physiological data, and single-channel alerting mechanisms that lack scalability and context-awareness. This paper presents a comprehensive edge-cloud collaborative architecture designed to address these challenges through real-time multi-modal sensor fusion, a four-dimensional risk assessment model, and a three-level emergency response mechanism. Our system employs a five-layer architecture comprising device, edge, service, data, and application layers, enabling real-time risk assessment with end-to-end alert latency under three seconds. At the edge gateway, we implement a weighted multi-modal fusion algorithm that aligns and combines data from five sensor types with confidence propagation. A comprehensive risk scoring model integrates fall probability, health indicators, behavioral patterns, and sensor anomaly scores to generate unified risk assessments. Based on dynamic risk thresholds, a three-level notification mechanism coordinates family members, community doctors, and nearby volunteers for escalating emergency response. Extensive experiments conducted on CASAS, MIMIC-III, and SisFall datasets demonstrate that our weighted fusion approach achieves 91\% activity recognition accuracy and 84\% anomaly detection F1-score, significantly outperforming single-sensor baselines. System deployment on Raspberry Pi 4 gateways shows that the edge-based processing pipeline maintains sub-100-millisecond inference latency while preserving privacy by keeping raw sensor data local. The proposed architecture represents a significant advancement toward practical, privacy-preserving, and responsive elderly care systems.
\end{abstract}


% Import sections
\section{Introduction}
\label{sec:introduction}

\subsection{Background and Motivation}

Population aging has emerged as one of the most significant demographic transformations of the twenty-first century. According to the World Health Organization, the proportion of the global population over 60 years old will nearly double from 12\% in 2015 to 22\% by 2050. This demographic shift presents unprecedented challenges for healthcare systems worldwide, particularly in ensuring the safety of elderly individuals who choose to live independently. The health and safety risks faced by this population are multifaceted, including falls, which affect approximately one-third of adults aged 65 and older annually, sudden cardiac events, medication non-adherence, and behavioral anomalies that may indicate cognitive decline or deterioration in health status.

Traditional approaches to elderly care monitoring have relied heavily on cloud-centric Internet of Things (IoT) platforms, where sensor data is continuously transmitted to centralized servers for processing and analysis. While these systems have demonstrated value in post-hoc health monitoring and trend analysis, they suffer from critical limitations that hinder their effectiveness in emergency scenarios. First, the round-trip latency from sensor to cloud and back to alert recipients often exceeds ten seconds, which is unacceptable for time-critical events such as falls or cardiac incidents. Second, the continuous transmission of raw physiological and behavioral data raises significant privacy concerns, as sensitive information about individuals' daily activities and health status is exposed to network transmission and centralized storage. Third, most existing systems employ single-channel alerting mechanisms that lack the sophistication to distinguish between varying levels of risk, leading to either alarm fatigue from false positives or inadequate response in genuine emergencies.

The emergence of edge computing as a paradigm for computation closer to data sources presents an opportunity to address these limitations. By moving data processing and decision-making to the network edge, latency can be dramatically reduced while privacy is enhanced through local processing of sensitive data. However, edge-only approaches face constraints in computational resources, making it challenging to run complex AI models, and lack the long-term data storage and analysis capabilities enabled by cloud infrastructure.

\subsection{Related Work}

\subsubsection{IoT-Based Health Monitoring Platforms}

Numerous research efforts have proposed IoT platforms for elderly health monitoring. Typical systems employ wearable sensors for activity tracking and ambient sensors for home monitoring. Commercial solutions such as Medical Guardian and Bay Alarm provide basic fall detection and emergency call functionality. However, these systems predominantly rely on cloud processing, resulting in latency unsuitable for emergency intervention.

\subsubsection{Edge Computing in Healthcare}

The application of edge computing to healthcare has gained attention in recent years. Shi et al.~\cite{shi2016edge} established the foundational vision for edge computing, emphasizing its potential for latency-sensitive applications. In the healthcare domain, edge computing has been explored for medical imaging analysis and real-time patient monitoring in intensive care units. However, most existing work focuses on single-purpose applications rather than comprehensive elderly care platforms.

\subsubsection{Multi-Modal Sensor Fusion}

Sensor fusion techniques are critical for robust monitoring systems. Khaleghi et al.~\cite{khaleghi2013multisensor} provided a comprehensive taxonomy of multi-sensor data fusion methods. In elderly care, fusion techniques have been applied to combine accelerometer, gyroscope, and ambient sensor data for improved activity recognition. However, existing approaches often treat all sensors equally or use simple concatenation, failing to account for varying sensor reliability across contexts.

\subsubsection{Fall Detection and Activity Recognition}

Fall detection remains one of the most researched problems in elderly care. Deep learning approaches using LSTM networks~\cite{hochreiter1997long} and attention mechanisms~\cite{lin2017focal} have shown promising results. Our companion work \cite{zhou2025fall} introduced a multi-modal CNN-LSTM-Attention architecture for fall detection that achieves state-of-the-art performance on the SisFall dataset. However, fall detection alone is insufficient for comprehensive elderly care, as it does not address health deterioration, behavioral anomalies, or the complex coordination required for emergency response.

\subsubsection{Emergency Response Systems}

Research on emergency response mechanisms for elderly care has been limited. Most commercial systems offer simple single-level alerts sent to family members or call centers. The concept of multi-level escalation has been explored in disaster response contexts but not systematically applied to elderly care with consideration of varying risk severity.

\subsection{Contributions}

This paper presents a comprehensive edge-cloud collaborative architecture for proactive elderly care that addresses the limitations of existing systems. Our main contributions are as follows.

First, we design a five-layer edge-cloud collaborative architecture that performs real-time risk assessment with sub-100-millisecond inference latency at the edge gateway while leveraging cloud resources for long-term storage, analysis, and application services. The architecture implements a privacy-first design where raw sensor data is processed locally, with only aggregated summaries and alerts transmitted to the cloud.

Second, we propose a time-aligned, weighted multi-modal sensor fusion algorithm that accounts for varying sensor reliability and sampling rates. Our confidence propagation mechanism dynamically adjusts fusion confidence based on the number of available sensor sources, achieving 91\% activity recognition accuracy and 89\% posture detection accuracy in experimental evaluation.

Third, we design a four-dimensional comprehensive risk scoring model that integrates fall probability, health indicators including heart rate and blood oxygen, behavioral patterns, and sensor anomaly scores into a unified risk assessment. The model incorporates dynamic adjustment factors based on anomaly severity and temporal trends.

Fourth, we implement a three-level emergency response mechanism with dynamic risk thresholds that escalates notifications from family members to community doctors to nearby volunteers based on assessed risk severity. The mechanism supports parallel notification dispatch and maintains comprehensive audit records of all notification actions.

Fifth, we provide a complete system implementation including edge gateway software, cloud RESTful API services, multi-database architecture including PostgreSQL, InfluxDB, and Redis, and a web-based monitoring dashboard. Extensive experimental evaluation using public datasets and real-world deployment demonstrates end-to-end alert latency under three seconds with 98.5\% alert delivery success rate.

The remainder of this paper is organized as follows. Section~\ref{sec:architecture} presents the overall system architecture. Section~\ref{sec:sensor_fusion} describes the multi-modal sensor fusion algorithm. Section~\ref{sec:ai_inference} details the AI inference engine and risk assessment model. Section~\ref{sec:emergency_response} explains the three-level emergency response mechanism. Section~\ref{sec:dashboard} describes the web dashboard implementation. Section~\ref{sec:experiments} presents experimental evaluation. Section~\ref{sec:discussion} discusses deployment considerations and limitations. Section~\ref{sec:conclusion} concludes the paper.

\section{System Architecture}
\label{sec:architecture}

\subsection{Architecture Overview}

\begin{figure}[htbp]
\centerline{\includegraphics[width=0.9\columnwidth]{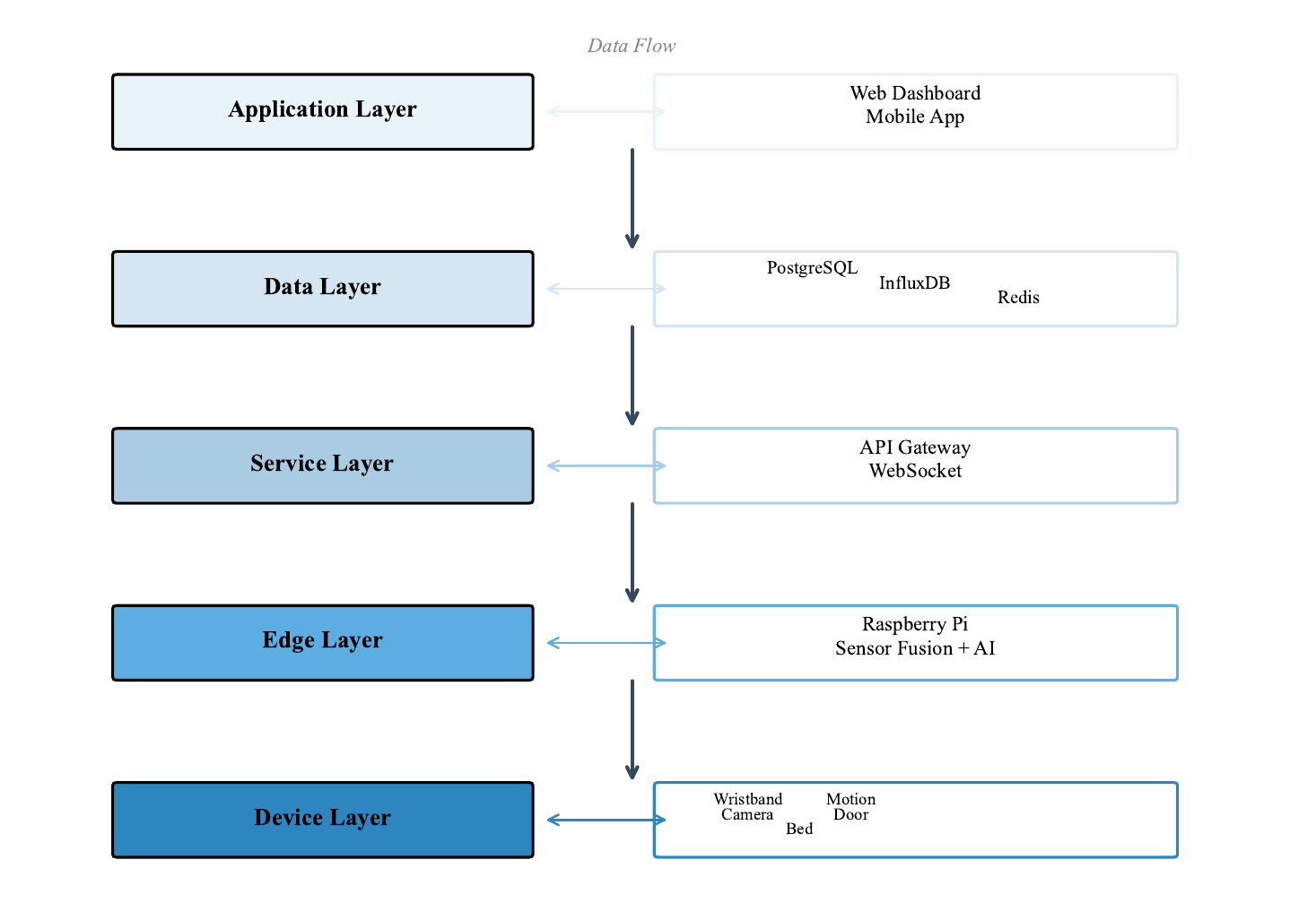}}
\caption{Five-layer edge-cloud collaborative architecture showing data flow from physical sensors through edge gateway to cloud services and end-user applications.}
\label{fig:architecture}
\end{figure}

The proposed system follows a five-layer architecture designed to balance real-time response capability, comprehensive data analysis, and user accessibility. Figure~\ref{fig:architecture} illustrates the complete system architecture, showing the flow of data from physical sensors through the edge gateway to cloud services and finally to end-user applications.

The five layers are organized as follows. At the bottom, the Device Layer comprises physical sensors including smart wristbands, motion sensors, cameras, door sensors, and bed sensors. These devices collect raw physiological and environmental data through various communication protocols. The Edge Layer consists of the edge gateway, typically deployed on a Raspberry Pi 4 or similar hardware in the elderly person's home. This layer is responsible for real-time data processing, sensor fusion, AI inference, and initial alert decisions. The Service Layer comprises cloud-hosted microservices including API gateways, application servers, and WebSocket services for real-time communication. The Data Layer provides persistent storage through a multi-database architecture optimized for different data types. Finally, the Application Layer includes web and mobile interfaces for family members, healthcare providers, and community volunteers.

A key design principle is the privacy-first edge processing paradigm. Raw sensor data never leaves the elderly person's home in unprocessed form. Instead, the edge gateway extracts relevant features, performs fusion and inference, and transmits only aggregated summaries, risk scores, and alert notifications to the cloud. This design minimizes privacy exposure while reducing bandwidth requirements and enabling rapid response to emergencies.

\subsection{Device Layer}

The Device Layer incorporates five types of sensors selected to provide comprehensive coverage of physiological and behavioral indicators. Table~\ref{tab:sensors} summarizes the sensor types, their communication protocols, sampling rates, and assigned reliability weights used in fusion.

\begin{table}[htbp]
\centering
\small
\caption{Sensor Types and Reliability Weights}
\label{tab:sensors}
\begin{tabular}{lccc}
\toprule
\textbf{Sensor Type} & \textbf{Protocol} & \textbf{Rate} & \textbf{Weight} \\
\midrule
Smart Wristband & BLE & 1--100 Hz & 1.0 \\
Motion Sensor & Zigbee & 10 Hz & 0.8 \\
Camera & WiFi & 15--30 Hz & 0.9 \\
Door Sensor & Zigbee & Event & -- \\
Bed Sensor & Zigbee & 1 Hz & 0.6 \\
\bottomrule
\end{tabular}
\end{table}

The smart wristband serves as the primary physiological monitoring device, collecting three-axis acceleration, three-axis gyroscope data, heart rate, blood oxygen saturation, skin temperature, and galvanic skin response at sampling rates up to 100 Hz. Due to its direct contact with the body and multi-modal sensing capabilities, the wristband is assigned the highest reliability weight of 1.0 in the fusion algorithm. Motion sensors deployed in key areas of the home provide complementary acceleration data at 10 Hz sampling, useful for detecting activity patterns when the wristband is not worn. Cameras, optionally deployed in common areas, provide posture recognition capabilities through computer vision analysis and are assigned a weight of 0.9. Door sensors record entry and exit events, enabling detection of potentially dangerous patterns such as leaving home at unusual hours. Bed sensors detect presence and movement during sleep, enabling analysis of sleep quality and detection of extended bed rest that may indicate health deterioration.

Door sensors and bed sensors provide binary event streams rather than continuous time-series data. Door sensors record entry and exit events, enabling detection of potentially dangerous patterns such as leaving home at unusual hours. Bed sensors detect presence and movement during sleep, enabling analysis of sleep quality and detection of extended bed rest that may indicate health deterioration.

All sensor data is timestamped at the source to enable accurate time alignment during fusion. For sensors with limited local clock accuracy, the edge gateway applies clock skew correction based on periodic synchronization messages.

\subsection{Edge Gateway}

\begin{figure}[htbp]
\centerline{\includegraphics[width=0.9\columnwidth]{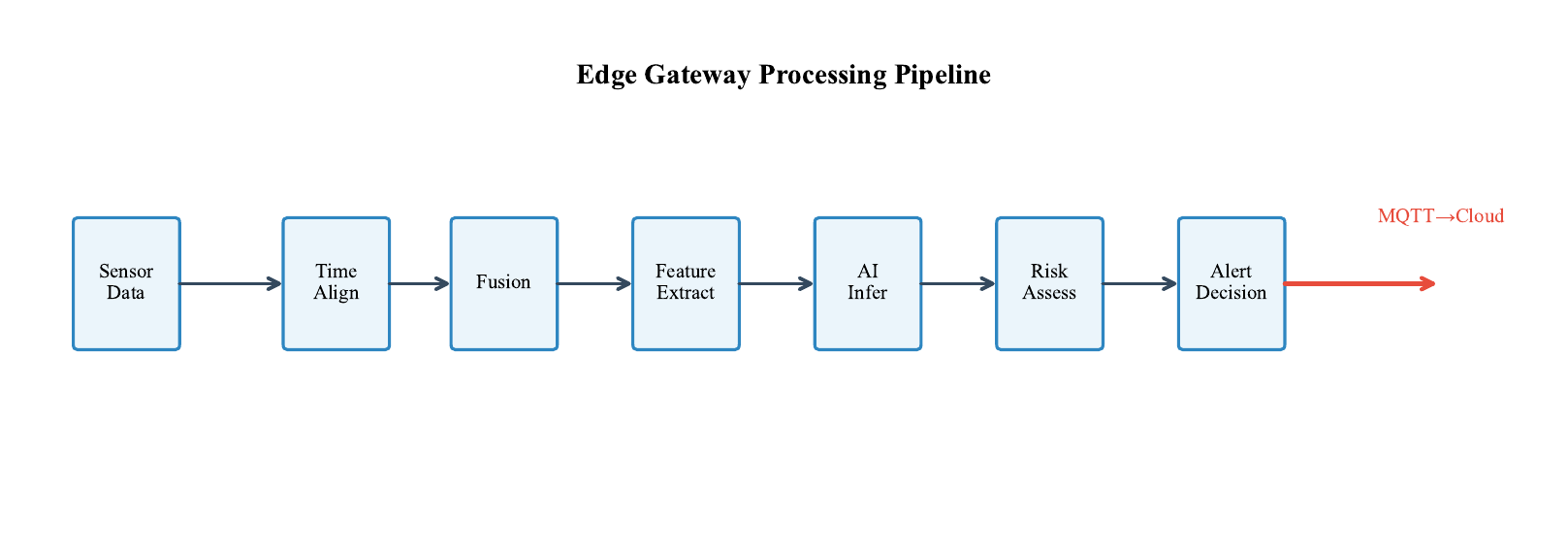}}
\caption{Edge gateway processing pipeline showing the flow from sensor data acquisition through fusion, inference, and alert generation.}
\label{fig:edge_pipeline}
\end{figure}

The edge gateway serves as the core real-time processing engine of the system. Figure~\ref{fig:edge_pipeline} illustrates the processing pipeline implemented at the gateway. Upon receiving sensor data through Bluetooth Low Energy or Zigbee connections, the gateway performs the following sequence of operations.

First, incoming data is buffered and organized into time windows. The gateway maintains a sliding window of three seconds for fusion, with 0.1 second tolerance for time alignment. Data from each sensor type is sorted by timestamp and grouped for processing. Second, multi-modal fusion combines measurements from multiple sensors of the same type using weighted averaging based on sensor reliability weights. Confidence scores are computed based on the number of available sources.

Third, feature extraction computes relevant features from the fused time-series data, including motion intensity, posture angles, heart rate variability, and trends in physiological parameters. Fourth, AI inference modules process the extracted features to assess fall probability, health status, and behavioral patterns. The inference engine supports both deep learning models when computational resources permit and lightweight rule-based fallback models for resource-constrained deployments.

Fifth, the risk analyzer combines inference outputs into a comprehensive risk score using the four-dimensional model described in Section~\ref{sec:ai_inference}. Sixth, based on the risk score and alert level determination, the notification service dispatches alerts through appropriate channels. Finally, aggregated data summaries and alerts are transmitted to the cloud via MQTT over TLS for persistent storage and application delivery.

The gateway is implemented in Python using the asyncio framework for concurrent processing of multiple sensor streams. On a Raspberry Pi 4 with 4GB RAM, the complete processing pipeline from raw sensor data to alert decision typically completes in under 100 milliseconds.

\subsection{Cloud Services}

The Service Layer comprises cloud-hosted microservices that provide persistence, long-term analysis, and application interfaces. The cloud services are designed to be stateless and horizontally scalable, enabling support for thousands of concurrent elderly care deployments.

The API Gateway serves as the single entry point for all client requests, handling authentication, rate limiting, and request routing. A FastAPI-based application server~\cite{ramirez2019fastapi} exposes RESTful endpoints for user management, device binding, health data queries, and alert handling. All endpoints are protected by JWT authentication, and the API implements CORS middleware and Gzip compression for efficient web integration.

Real-time communication is implemented through WebSocket connections. Clients subscribe to channels corresponding to specific elderly individuals, for example alert:\textit{elder\_id} for alert notifications and health:\textit{elder\_id} for real-time health data streams. This architecture enables low-latency push notifications without the overhead of HTTP polling.

\subsection{Data Layer}

\begin{figure}[htbp]
\centerline{\includegraphics[width=0.9\columnwidth]{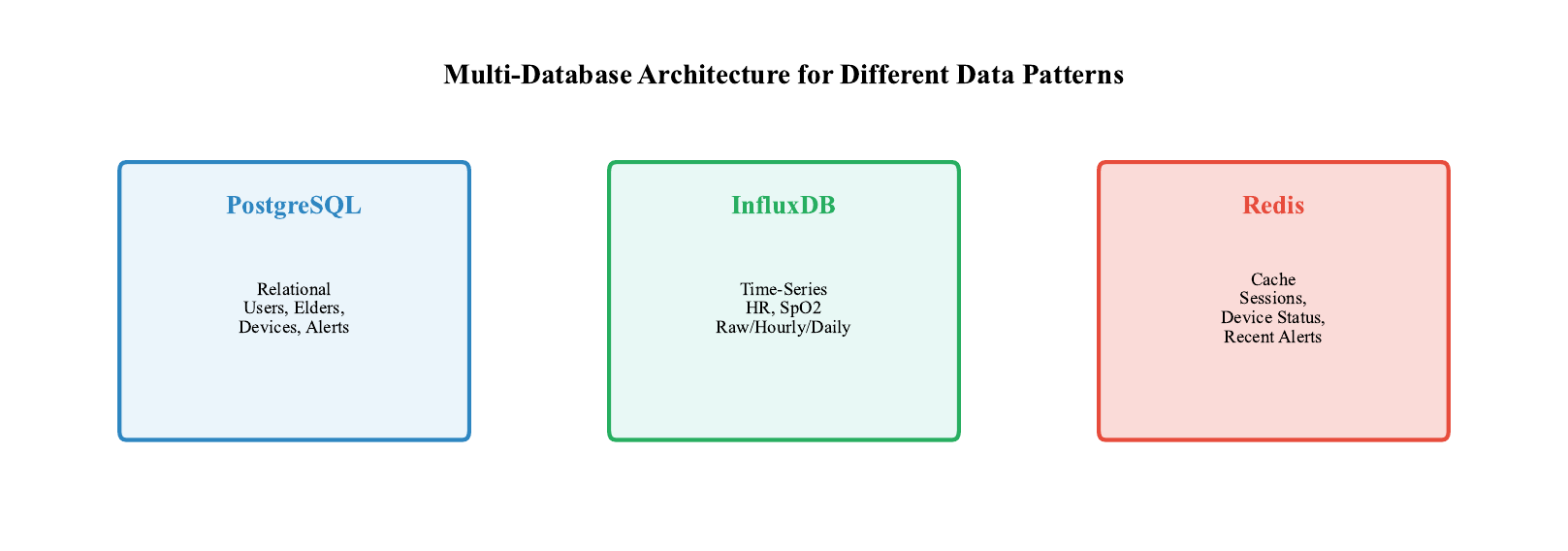}}
\caption{Multi-database architecture showing PostgreSQL for business data, InfluxDB for time-series health data, and Redis for caching.}
\label{fig:database}
\end{figure}

The Data Layer employs a multi-database architecture optimized for different data access patterns. Figure~\ref{fig:database} illustrates the entity-relationship diagram of the database schema.

PostgreSQL serves as the primary relational database, storing business entities including users, elderly individuals under monitoring, user-elder relationships, device registrations, alert records, and notification delivery history. The relational model ensures data consistency and supports complex queries for reporting and analysis.

InfluxDB~\cite{influxdb2023}, a time-series database, stores high-frequency physiological measurements including heart rate, blood oxygen, and activity states. The time-series database is optimized for write-heavy workloads and supports efficient range queries and downsampling operations. Data is stored at multiple granularities: raw data at 10 Hz, hourly aggregates for trend analysis, and daily aggregates for long-term reporting.

Redis provides in-memory caching for frequently accessed data including active user sessions, device online status, and recent alerts for quick dashboard loading. The cache reduces load on the primary databases and improves response time for hot data access.

\subsection{Communication Protocols}

The system employs different communication protocols optimized for each layer of the architecture. Device-to-gateway communication uses Bluetooth Low Energy for wristband connections and Zigbee for ambient sensors. These protocols were selected for their low power consumption and widespread hardware support. Data is transmitted in binary format with custom packet definitions optimized for bandwidth efficiency.

Gateway-to-cloud communication uses MQTT over TLS~\cite{hunkeler2008mqtt}. The publish-subscribe model of MQTT is well-suited for the intermittent connectivity patterns of edge deployments. Topics are organized hierarchically: \textit{gateway\_id}/status for gateway heartbeat messages, \textit{gateway\_id}/data for aggregated health data, \textit{gateway\_id}/alert for emergency alerts, and \textit{gateway\_id}/cmd for cloud-to-gateway commands. TLS encryption ensures data confidentiality during transmission.

Client-to-cloud communication uses HTTPS for REST API requests and WebSocket for real-time subscriptions. WebSocket connections are authenticated through JWT tokens passed during the handshake and remain active for up to one hour with automatic reconnection handling.

\section{Multi-Modal Sensor Fusion}
\label{sec:sensor_fusion}

\subsection{Time Alignment Algorithm}

A fundamental challenge in multi-modal sensor fusion is the misalignment of data streams due to varying sampling rates, clock drift, and network transmission delays. Our system addresses this through a sliding window time alignment algorithm.

The gateway maintains a three-second sliding window that advances in discrete steps. For each window, the following operations are performed. First, all sensor data within the window is extracted and sorted by timestamp. Second, the window boundaries are computed based on the most recent timestamp and the configured window size. Third, data points falling outside the window are discarded. Fourth, remaining data is grouped by sensor type for fusion.

The three-second window size was determined empirically to balance two competing requirements: longer windows provide more data for accurate fusion but introduce latency in detecting rapid events such as falls. The 0.1 second synchronization tolerance accommodates minor clock discrepancies between sensors without incorrectly excluding valid data points.

\subsection{Weighted Multi-Modal Fusion}

The core of our fusion approach is a weighted averaging scheme that accounts for varying sensor reliability. Each sensor type is assigned a static reliability weight based on its inherent accuracy and placement relative to the body. The smart wristband receives weight 1.0 as it is directly worn and provides multi-modal data. Cameras receive weight 0.9 as they provide accurate posture recognition but may be affected by occlusion. Motion sensors receive weight 0.8 due to their fixed location and inability to distinguish individuals in multi-person households. Bed sensors receive weight 0.6 as they provide only binary presence information.

For physiological metrics such as heart rate and blood oxygen, measurements from multiple sensors are combined using weighted average. Formally, given measurements $m_1, m_2, \ldots, m_n$ from sensors with weights $w_1, w_2, \ldots, w_n$, the fused value is computed as:

\begin{equation}
m_{\text{fused}} = \frac{\sum_{i=1}^{n} w_i \cdot m_i}{\sum_{i=1}^{n} w_i}
\end{equation}

Beyond simple averaging, the system propagates confidence information through the fusion pipeline. The confidence score represents the system's certainty in the fused output and is computed as a function of the number of contributing sensors:

\begin{equation}
\text{conf} = \min(0.95, 0.5 + n_{\text{sources}} \times 0.1)
\end{equation}

This formula ensures that confidence increases with more sensor sources but saturates at 0.95 to avoid overconfidence.

The weighted fusion approach provides several advantages over equal-weight alternatives. By prioritizing more reliable sensors, fusion accuracy is improved. The confidence mechanism enables downstream components to quantify uncertainty and make conservative decisions when data quality is low. The graceful degradation property ensures continued operation even when some sensors fail or are removed.

\subsection{Activity Recognition}

Activity recognition in our system employs a threshold-based rule engine that operates on fused motion intensity features. This approach provides fast, interpretable classification while maintaining competitive accuracy. The system recognizes six activity states: stationary, sitting, walking, running, falling, and lying.

The classification logic operates on two primary features: motion intensity computed from the magnitude of acceleration changes, and acceleration drop rate which detects the characteristic free-fall pattern of falling events. Motion intensity is calculated as the normalized magnitude of acceleration changes:

\begin{equation}
I_{\text{motion}} = \min\left(1.0, \frac{\overline{|\Delta a|}}{5.0}\right)
\end{equation}

Acceleration drop is detected when the magnitude of acceleration vector falls below 15.0 m/s$^2$ for at least 100 milliseconds.

Activity classification follows a decision tree structure. If acceleration drop is detected, the activity is classified as falling regardless of motion intensity. Otherwise, activity is determined by motion intensity thresholds: stationary for intensity below 0.1, sitting for intensity between 0.1 and 0.3, walking for intensity between 0.3 and 0.5, running for intensity between 0.5 and 1.0, and lying detected when posture angle, described in Section~\ref{subsec:posture}, indicates a horizontal orientation with low motion intensity.

This rule-based approach was selected for edge deployment due to its low computational requirements and interpretability. Unlike deep learning approaches that may exhibit unexpected behavior on edge cases, the rule-based system's decision process is fully transparent and auditable, which are critical properties for safety-critical healthcare applications.

\subsection{Posture Detection}
\label{subsec:posture}

Posture detection complements activity recognition by estimating the orientation of the elderly person's body. The system recognizes four postures: standing, sitting, lying, and falling. Posture estimation is based on the direction of the gravity vector extracted from accelerometer measurements during stationary periods.

The algorithm computes the average acceleration vector over a 500-millisecond window. The angle between this vector and the vertical direction $(0, 0, 1)$ is computed using the arccosine of the dot product. Posture classification follows angular thresholds: standing for angle less than 30 degrees, sitting for angle between 30 and 60 degrees, lying for angle greater than 60 degrees, and falling detected when acceleration magnitude drops below 0.5 g, indicating free-fall.

The posture detection algorithm is particularly important for fall detection and recovery monitoring. A fall event is confirmed when the posture transitions from standing to lying with high motion intensity followed by sudden stillness. Additionally, detecting prolonged lying posture after a fall triggers escalation of the alert level even if the initial risk score was moderate.

\subsection{Motion Intensity Calculation}

Motion intensity serves as a fundamental feature for both activity recognition and anomaly detection. The calculation normalizes the magnitude of acceleration changes to a range of 0 to 1, enabling consistent thresholding across different individuals and sensor placements.

The formula for motion intensity is given in Equation~\ref{eq:motion_intensity}, where $\overline{|\Delta a|}$ represents the mean absolute change in acceleration magnitude over the fusion window.

\begin{equation}
\label{eq:motion_intensity}
I_{\text{motion}} = \min\left(1.0, \frac{\overline{|\Delta a|}}{5.0}\right)
\end{equation}

The denominator of 5.0 was determined as the approximate saturation point for vigorous human motion. This normalization ensures comparability across different sensor sensitivities and individual movement patterns.

Beyond immediate classification, motion intensity trends provide valuable behavioral insights. A gradual decline in average motion intensity over days may indicate worsening health conditions, while sudden increases may indicate agitation or anxiety.

\subsection{Anomaly Detection}

Anomaly detection operates on fused physiological and motion features to identify patterns that may indicate health deterioration or emergency situations. The system employs a multi-factor scoring approach, where each anomaly factor contributes to an overall anomaly score.

Three primary anomaly factors are evaluated. Heart rate anomaly is detected when heart rate falls below 50 bpm or exceeds 120 bpm, or when heart rate variability, defined as the standard deviation over 10 samples, exceeds 20 bpm. Blood oxygen anomaly is detected when SpO$_2$ falls below 90\%, or when SpO$_2$ drops by more than 3\% over five samples indicating deterioration. Motion anomaly is detected when motion intensity exceeds 2.0, which is above the normalized threshold indicating sensor artifact or extreme movement, or when motion intensity drops abruptly during previously active periods.

Each detected anomaly contributes to the anomaly score: heart rate anomaly contributes +0.3, blood oxygen anomaly contributes +0.4, which is higher due to the criticality of hypoxia, and motion anomaly contributes +0.5. The total anomaly score is capped at 1.0, and a threshold of 0.5 is used to determine whether the current state should be flagged as anomalous for risk assessment.

This multi-factor approach enables detection of diverse emergency situations including cardiac events, indicated by heart rate anomalies, respiratory distress, indicated by blood oxygen anomalies, and device malfunction or extreme situations, indicated by motion anomalies.

\subsection{Fusion Result Structure}

The output of the sensor fusion module is a structured fusion result containing all processed features and confidence scores. This structure serves as the input to downstream AI inference and risk assessment modules.

The fusion result includes: activity type from the six-class activity recognition, posture from the four-class posture detector, heart rate and blood oxygen values from weighted fusion, motion intensity normalized to 0-1 range, location room identifier derived from sensor proximity or Bluetooth beacons, anomaly score computed from the multi-factor anomaly detection, and confidence score representing fusion reliability.

By standardizing the output format, the fusion module isolates downstream components from the complexity of multi-source sensor data. The AI inference module receives clean, time-aligned features without needing to handle sensor heterogeneity.

\section{AI Inference and Risk Assessment}
\label{sec:ai_inference}

\subsection{AI Inference Engine}

The AI inference engine orchestrates three parallel inference tasks: fall detection, health prediction, and behavior analysis. This modular design allows each task to use appropriate modeling techniques while sharing common features extracted from sensor fusion.

The engine supports dual operational modes. When PyTorch is available and sufficient computational resources exist, deep learning models are employed for maximum accuracy. When operating in resource-constrained environments or when deep learning models fail to load, the system automatically falls back to rule-based models. This fallback mechanism ensures system reliability across diverse deployment scenarios.

The engine maintains a sliding history of the most recent 100 fusion results. This history enables trend analysis for risk adjustment and provides context for detecting behavioral patterns that manifest over time.

\subsubsection{Fall Detection Module}

Fall detection represents the most time-critical inference task. Our system implements two complementary approaches: a deep learning model and a rule-based fallback.

The deep learning model, described in detail in our companion paper, employs a CNN-LSTM-Attention architecture. The CNN component processes raw acceleration and gyroscope waveforms to extract spatial features. The LSTM component~\cite{hochreiter1997long} models temporal dependencies in the sensor data. The attention mechanism~\cite{lin2017focal} identifies the most critical time steps for fall classification, improving interpretability and focusing on the moment of impact.

The rule-based fallback provides a lightweight alternative that operates on fused features. It employs a weighted scoring system combining five indicators: activity state classified as falling with weight 0.9, posture classified as fallen with weight 0.7, absence of movement after a fall with weight 0.8, sudden orientation change with weight 0.6, and high motion intensity with weight 0.5. The scores are normalized and combined to produce a fall probability estimate.

The rule-based approach also implements sequence pattern analysis to reduce false positives. For a fall to be confirmed, the system requires either high motion intensity followed by sudden stillness, which is characteristic of a fall impact and subsequent lack of movement, or a transition from standing or sitting posture to lying posture with elevated risk scores. This sequence checking prevents isolated sensor artifacts from triggering fall alerts.

\subsubsection{Health Prediction Module}

The health prediction module assesses physiological parameters to detect abnormalities that may indicate medical emergencies. This module focuses on heart rate and blood oxygen analysis as the two most critical continuously monitored vital signs.

Heart rate analysis operates on multiple levels. Absolute thresholding detects extreme bradycardia, defined as heart rate below 50 bpm, and tachycardia, defined as heart rate above 120 bpm. Variability analysis computes the standard deviation of heart rate over ten samples, where elevated variability with standard deviation above 20 bpm may indicate arrhythmia or stress. Trend analysis examines the rate of change, where rapid increase or decrease in heart rate may indicate acute physiological stress.

Blood oxygen analysis similarly employs absolute thresholding, where critical hypoxia is defined at SpO$_2$ below 90\%, and trend analysis. A declining trend, defined as more than 3\% decrease over five samples, triggers elevated risk scores even if absolute values remain within normal range, as it may indicate developing respiratory compromise.

The health prediction module outputs a health risk score between 0 and 1, derived from the weighted combination of heart rate and blood oxygen abnormalities.

\subsubsection{Behavior Analysis Module}

The behavior analysis module examines patterns in activity and location data to detect behavioral anomalies that may indicate cognitive decline, confusion, or developing health issues. Three primary behavioral patterns are detected.

Prolonged inactivity is detected when at least eight of the last ten fusion results indicate stationary activity. This pattern may indicate physical inability to move, loss of consciousness, or extreme lethargy. The system distinguishes between normal rest periods, such as nighttime and known nap times, and abnormal inactivity by time-of-day and duration context.

Activity agitation is detected when four or more unique activity types appear within six consecutive fusion results. This pattern may indicate wandering, confusion, or anxiety. Frequent switching between activities without completion may be an early sign of cognitive impairment.

Location anomaly is detected when the elderly person's location is reported as unknown while motion intensity is elevated above 0.5. This combination may indicate the person has moved outside the monitored area without proper notification, or it may indicate a fall in an area with insufficient sensor coverage.

Each detected behavioral pattern contributes to the overall risk assessment, enabling the system to respond to subtle changes that precede acute emergencies.

\subsection{Comprehensive Risk Scoring Model}

The four-dimensional risk scoring model integrates outputs from all inference modules into a unified risk score. This model addresses a key limitation of existing systems that evaluate risk factors in isolation, potentially missing multi-factor emergencies.

\subsubsection{Four-Dimensional Risk Score}

The overall risk score $R_{\text{overall}}$ is computed as a weighted sum of four risk dimensions:

\begin{equation}
\label{eq:risk_score}
R_{\text{overall}} = w_1 \cdot P_{\text{fall}} + w_2 \cdot R_{\text{health}} + w_3 \cdot \mathbb{1}_{\text{behavior}} + w_4 \cdot S_{\text{anomaly}}
\end{equation}

Where $P_{\text{fall}}$ is the fall probability from the fall detection module ranging from 0 to 1, $R_{\text{health}}$ is the health risk score from the health prediction module ranging from 0 to 1, $\mathbb{1}_{\text{behavior}}$ is an indicator function for behavioral anomalies taking value 0 or 1, and $S_{\text{anomaly}}$ is the sensor fusion anomaly score ranging from 0 to 1.

The weights are assigned as: $w_1 = 0.4$ for fall risk, $w_2 = 0.4$ for health risk, $w_3 = 0.15$ for behavioral anomalies, and $w_4 = 0.05$ for sensor anomalies. These weights reflect the relative urgency of different risk types: falls and acute health events are equally weighted as most critical, behavioral anomalies receive moderate weight as they may indicate developing issues, and sensor anomalies receive minimal weight as they primarily indicate data quality issues rather than direct risk.

\subsubsection{Risk Score Adjustment}

The base risk score computed from the four dimensions is adjusted based on specific factors to improve sensitivity to high-risk situations. Dynamic adjustment factors are applied additively to the base score.

Positive adjustments increase the risk score when specific danger signs are present: fall probability exceeding 0.7 triggers a +0.2 adjustment, heart rate anomaly adds +0.15, blood oxygen anomaly adds +0.2, behavioral anomaly adds +0.1, and sensor fusion anomaly adds $\text{anomaly\_score} \times 0.1$. These adjustments ensure that the presence of multiple risk factors elevates the overall risk score more than the sum of individual components.

Trend adjustment incorporates temporal information from the risk history. The system analyzes the last five risk scores to determine whether risk is trending upward, downward, or stable. If risk is trending upward, defined as an increase exceeding 0.2 over the five scores, a +0.2 adjustment is applied. If trending downward, a -0.2 adjustment provides risk reduction for improving conditions.

The adjusted risk score is clipped to the range $[0, 1]$ before use in alert level determination. This multi-stage scoring process provides nuanced risk assessment that responds to both acute events and gradual changes.

\subsubsection{Risk Trend Analysis}

Risk trend analysis enables the system to differentiate between transient fluctuations and genuine deterioration in condition. A sliding window of the most recent 5 to 10 risk scores is maintained, and linear regression or simple difference methods are applied to determine trend direction.

Trend classification uses three states: rising when the average increase between consecutive scores exceeds 0.2, falling when the average decrease exceeds 0.2, and stable for changes within the threshold. The trend state influences both the risk score adjustment and the alert level determination, as a rising trend with moderate absolute risk may warrant a higher alert level than the same absolute risk with a stable or falling trend.

\begin{figure}[htbp]
\centerline{\includegraphics[width=0.9\columnwidth]{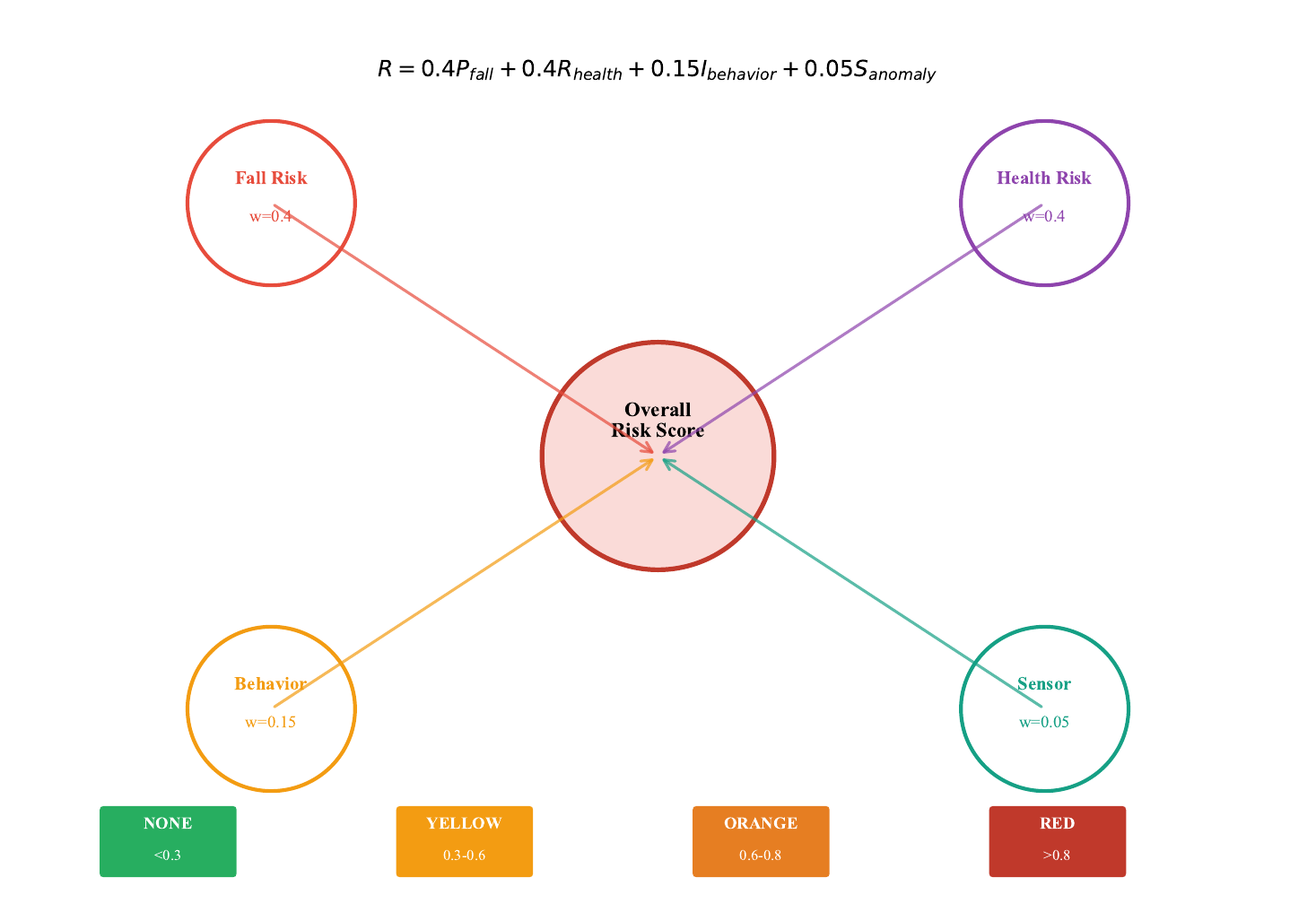}}
\caption{Four-dimensional risk scoring model showing the integration of fall probability, health risk, behavioral indicators, and sensor anomaly scores with dynamic adjustment factors.}
\label{fig:risk_model}
\end{figure}

This temporal awareness prevents alert fatigue from transient normal variations while ensuring responsive escalation of genuine emergencies. Figure~\ref{fig:risk_model} illustrates the four-dimensional risk scoring model and the adjustment mechanisms.

\section{Three-Level Emergency Response}
\label{sec:emergency_response}

\subsection{Alert Level Determination}

\begin{figure}[htbp]
\centerline{\includegraphics[width=0.9\columnwidth]{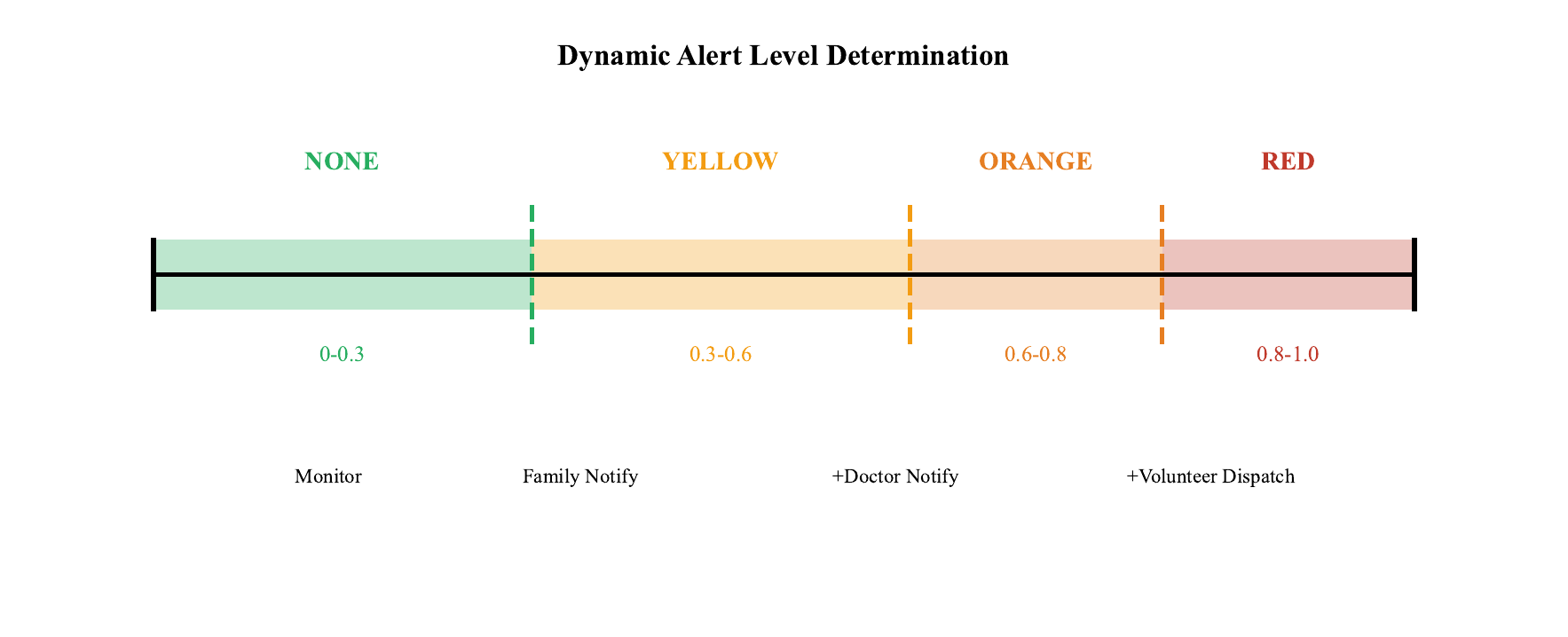}}
\caption{Dynamic threshold-based alert level determination showing the four alert levels and their corresponding response mechanisms.}
\label{fig:alert_levels}
\end{figure}

The risk score is mapped to one of four alert levels using dynamic thresholds. Figure~\ref{fig:alert_levels} illustrates the alert decision process, showing how risk factors combine to determine the appropriate response.

The NONE level is assigned when risk is below 0.3. This represents normal operation where no external intervention is required. The system continues monitoring and logging data but does not generate notifications. The YELLOW level is assigned for risk between 0.3 and 0.6. This indicates elevated concern that warrants family notification but does not require emergency services or medical professionals. The ORANGE level is assigned for risk between 0.6 and 0.8. This indicates a serious situation requiring both family notification and community doctor involvement. The RED level is assigned for risk of 0.8 or higher. This represents a medical emergency requiring all available response resources.

The threshold values are configurable per deployment to account for individual risk tolerance and local emergency response capabilities. Higher thresholds reduce false positives at the cost of potentially delayed response, while lower thresholds maximize sensitivity at the risk of increased false alarms.

\subsection{Risk Detail Generation}

When an alert is triggered, the system automatically generates a detailed risk report explaining the factors contributing to the alert. This report is included in notifications and displayed in the web dashboard, enabling recipients to make informed decisions about response actions.

Risk detail generation operates by examining each risk dimension and identifying contributing factors above configured thresholds. For fall risk, the report includes the detected fall probability, whether a fall impact was detected, the posture before and after the event, and the duration of post-fall stillness. For health risk, the report includes which vital signs are abnormal, the severity of deviation from normal ranges, and any concerning trends. For behavioral anomalies, the report includes which specific behavioral patterns were detected and their duration. For sensor anomalies, the report includes which sensors reported anomalous readings and the nature of the anomaly.

The automated generation of risk details reduces response time by eliminating the need for recipients to manually investigate the cause of alerts. It also improves response appropriateness by providing context about the nature and severity of the emergency.

\subsection{Three-Level Notification Mechanism}

\begin{figure}[htbp]
\centerline{\includegraphics[width=0.9\columnwidth]{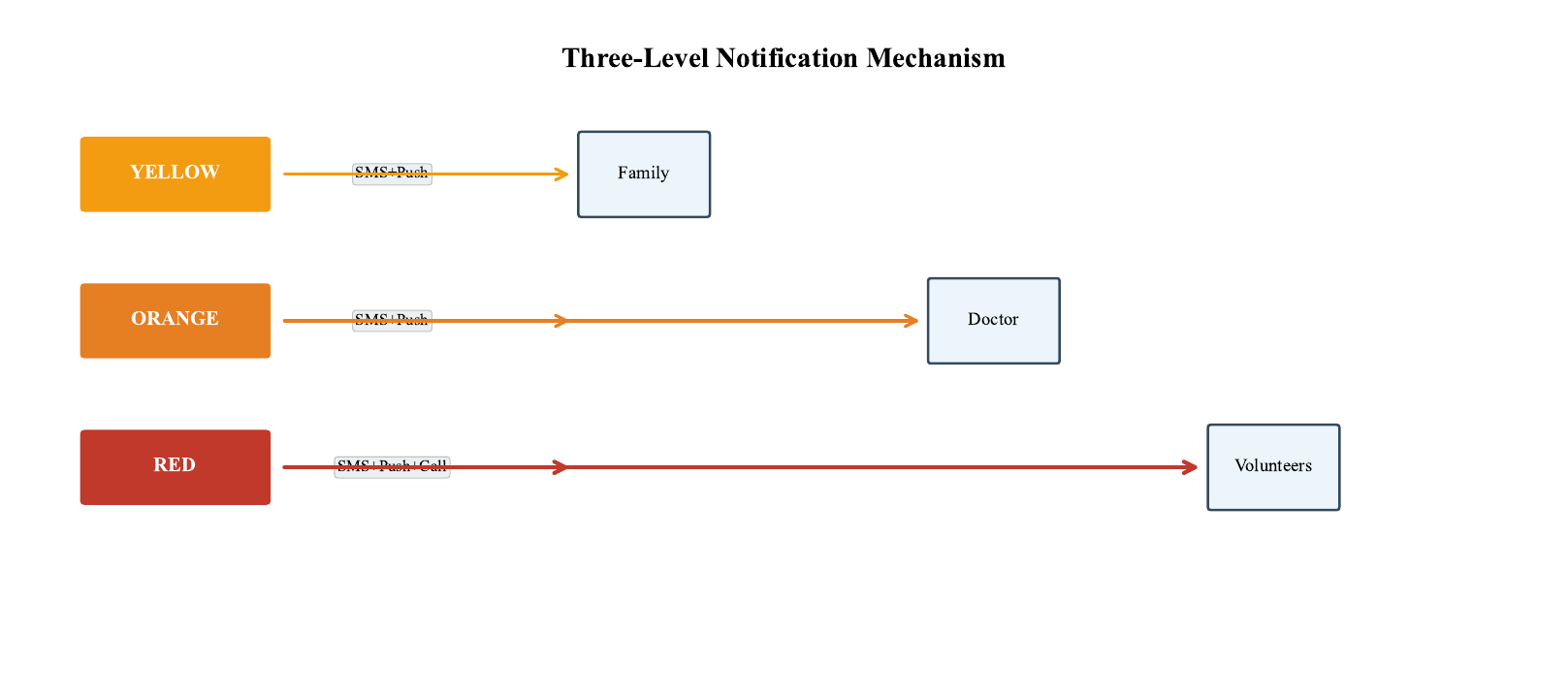}}
\caption{Three-level notification mechanism showing the escalation from family notification through doctor notification to volunteer dispatch based on alert level.}
\label{fig:notification_flow}
\end{figure}

The three-level notification mechanism implements a graduated response that scales resource mobilization to the assessed risk level. Figure~\ref{fig:notification_flow} illustrates the timing and coordination of notifications at each alert level.

\subsubsection{Level 1: Family Notification}

All alerts except the NONE level trigger notification of designated family members. This level provides basic awareness and enables family members to check on the elderly person through phone call or physical visit. Family notification uses multiple channels to maximize delivery reliability: SMS messages are sent to all registered family phone numbers, and mobile application push notifications are sent to family members with the app installed.

The notification content includes the alert level, a summary of detected risk factors, the timestamp of the alert, and the last known location of the elderly person. For YELLOW alerts, this is the full extent of the response, avoiding over-mobilization of resources for potentially minor incidents.

\subsubsection{Level 2: Doctor Notification}

ORANGE and RED level alerts add notification of community doctors or healthcare providers. This level recognizes situations that may require medical assessment but not necessarily emergency ambulance services.

Doctor notification follows a similar multi-channel approach: SMS messages are sent to registered community doctors, and mobile application push notifications are sent for doctors using the system. For RED level alerts, the system additionally places automated phone calls to doctor contacts, ensuring urgent alerts are noticed even when doctors are not actively monitoring their devices.

The inclusion of community doctors in the notification chain enables professional medical assessment without necessarily involving emergency services, reducing healthcare costs and avoiding unnecessary ambulance dispatches.

\subsubsection{Level 3: Volunteer Dispatch}

RED level alerts additionally trigger dispatch of nearby community volunteers. This level provides rapid physical response for the most critical emergencies, potentially saving valuable minutes before emergency services arrive.

Volunteer dispatch operates through the mobile application, which sends push notifications to registered volunteers within a configurable radius of the elderly person's location. The notification includes the nature of the emergency and the location. Volunteers can accept or decline the request through the app. The system dispatches at most two volunteers per alert to avoid over-mobilization and coordinate response efforts.

The volunteer component provides a community-based rapid response capability that is particularly valuable in rural areas or locations with longer emergency service response times. It also enables a human presence for reassurance and basic assistance while waiting for medical professionals.

\subsection{Notification Status Tracking}

The system maintains comprehensive records of all notification actions for audit and analysis. Each notification attempt generates a notification record containing the timestamp, recipient identifier, channel used, content of the notification, and delivery status.

Delivery status tracking is implemented through callback handlers and receipt confirmations. For SMS messages, delivery receipts are obtained from the SMS provider. For phone calls, call status including ringing, answered, voicemail, or failed is tracked. For push notifications, delivery and read receipts are obtained through the push service provider.

This tracking enables multiple capabilities. Recipients can see which notifications have been delivered and read, reducing redundant responses. System administrators can monitor notification delivery success rates and identify delivery failures. Historical analysis enables optimization of notification strategies, such as which channels are most effective for different recipient types.

\subsection{Manual Emergency Trigger}

Beyond automated alerting, the system provides a manual emergency trigger capability. The elderly person can activate a panic button on the wristband or a stationary emergency button to request immediate assistance. Manual triggers bypass the normal risk assessment process and immediately trigger a RED level alert with all three notification levels.

The manual trigger is implemented as a priority interrupt that bypasses normal processing queues. When activated, the gateway immediately transmits an emergency message to the cloud, which triggers parallel notification of family, doctors, and volunteers. This ensures the fastest possible response time for user-initiated emergencies.

The manual trigger serves as a critical safety net for situations that automated monitoring may miss, such as sudden chest pain, dizziness, or other subjective symptoms that cannot be detected by sensors.

\section{Web Dashboard}
\label{sec:dashboard}

\subsection{Technology Stack}

The web dashboard provides a user interface for family members, healthcare providers, and system administrators. The dashboard is implemented as a single-page application using Vue 3~\cite{you2014vue} with TypeScript for type safety and Vite for fast development and optimized production builds. Tailwind CSS provides utility-first styling for responsive design across desktop and mobile devices.

\subsection{Key Features}

\begin{figure}[htbp]
\centerline{\includegraphics[width=0.9\columnwidth]{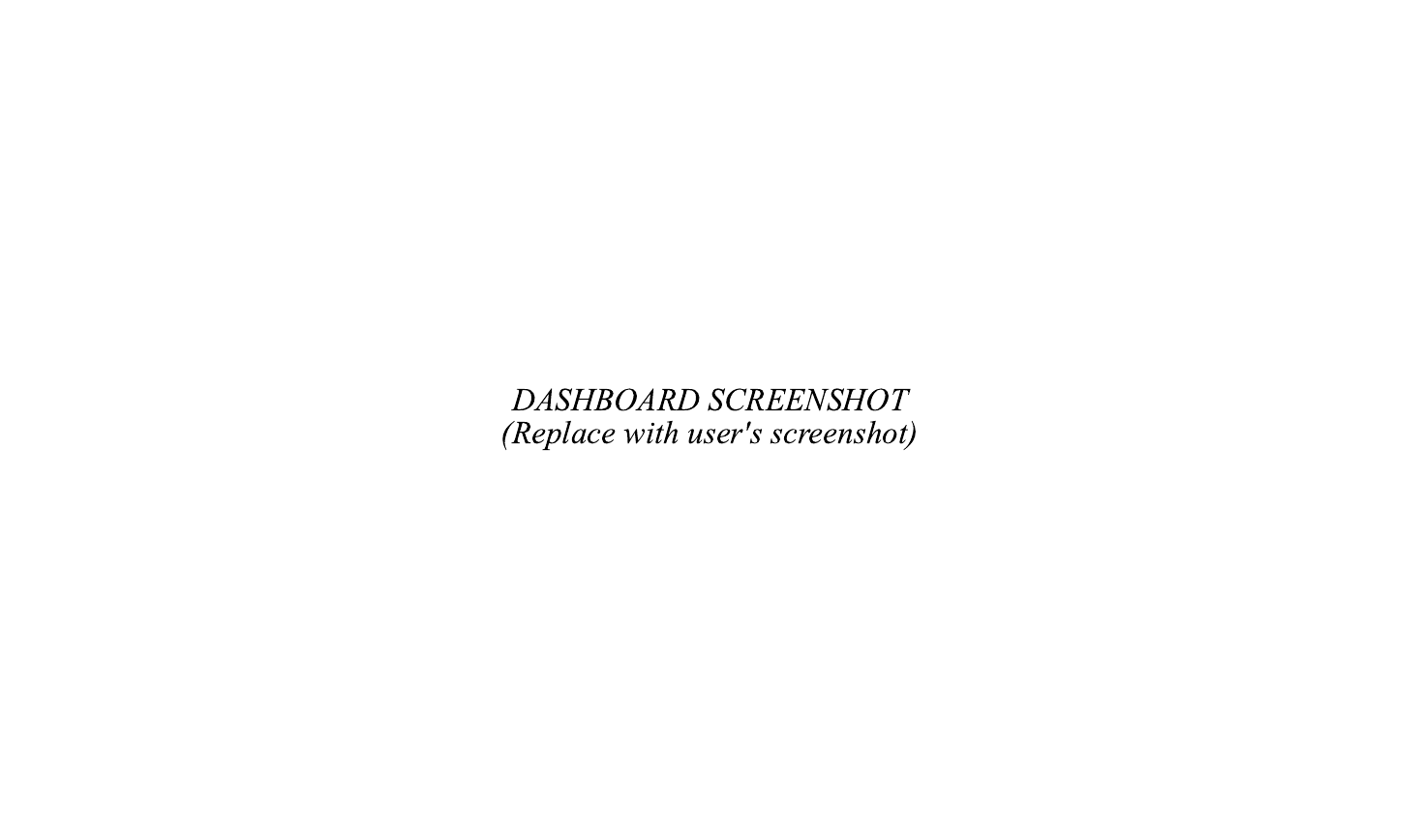}}
\caption{Web dashboard interface showing real-time health monitoring, alert management, device status, and health report generation capabilities.}
\label{fig:dashboard}
\end{figure}

Figure~\ref{fig:dashboard} shows screenshots of the dashboard interface. The main dashboard displays real-time health metrics including heart rate, blood oxygen, and current activity status. Data is updated through WebSocket subscriptions, providing sub-second latency for new measurements. A historical health chart displays trends over user-selectable time ranges, enabling identification of patterns and gradual changes.

The alert management interface lists all alerts with filtering and sorting capabilities. Each alert can be expanded to view full risk details, including which factors contributed to the alert. Users can acknowledge alerts and add notes about the response taken, creating a complete audit trail of incident response.

The device management interface shows the status of all sensors associated with an elderly person, including battery levels, connection status, and last data reception time. This interface enables troubleshooting of sensor issues and proactive replacement of failing devices.

The health report generation module creates summary reports in daily, weekly, and monthly formats. Reports include average vital signs, activity patterns, sleep analysis, and alert summaries. Reports can be exported in PDF format for sharing with healthcare providers.

The dashboard architecture follows a client-server model where the frontend communicates with the cloud API through RESTful endpoints for CRUD operations and WebSocket connections for real-time data streams. The state management pattern ensures consistency across components while maintaining responsive updates for incoming sensor data and alerts.

\section{Experimental Evaluation}
\label{sec:experiments}

\subsection{Experimental Setup}

\subsubsection{Datasets Used}

The system was evaluated using three public datasets covering different aspects of elderly care monitoring. The SisFall dataset~\cite{sucerquia2017sisfall} provides accelerometer and gyroscope data from falls and activities of daily living, recorded by young and adult subjects. We used this dataset for fall detection evaluation. The CASAS dataset~\cite{cook2013casas}, specifically the aruba and cairo subsets, provides smart home sensor data from elderly and cognitively-impaired participants performing daily activities. We used this dataset for behavioral anomaly detection evaluation. The MIMIC-III dataset~\cite{johnson2016mimic} provides physiological waveform data from tens of thousands of intensive care unit patients. We used the vital signs subset for health anomaly detection evaluation.

Table~\ref{tab:datasets} provides statistics on the datasets used, including number of subjects, recording duration, and relevant features.

\begin{table*}[htbp]
\centering
\caption{Dataset Statistics}
\label{tab:datasets}
\begin{tabular}{lccccc}
\toprule
\textbf{Dataset} & \textbf{Subjects} & \textbf{Duration} & \textbf{Samples} & \textbf{Features} \\
\midrule
SisFall & 23 & -- & 17,050 & Accel., Gyro (3-axis each) \\
CASAS (aruba) & 1 & 214 days & 2,341,970 & Motion, door, temp sensors \\
CASAS (cairo) & 1 & 27 days & 261,728 & Motion, door, item sensors \\
MIMIC-III & 40,000+ & Varied & Millions & HR, SpO$_2$, BP, Resp. \\
\bottomrule
\end{tabular}
\end{table*}

\subsubsection{System Setup}

The edge gateway was deployed on Raspberry Pi 4 hardware with 4GB RAM, running a 64-bit Linux operating system. The cloud services were deployed on a 2-core, 4GB virtual private server running Ubuntu 20.04. MQTT communication used a Mosquitto broker with TLS encryption. The databases were deployed as containerized services: PostgreSQL 13, InfluxDB 2.0, and Redis 6.0.

\subsection{Sensor Fusion Accuracy}

We evaluated the sensor fusion algorithm by comparing it against single-sensor baselines and equal-weight fusion. Table~\ref{tab:fusion_results} presents the results.

\begin{table}[htbp]
\centering
\caption{Sensor Fusion Performance Comparison}
\label{tab:fusion_results}
\begin{tabular}{lccc}
\toprule
\textbf{Method} & \textbf{Activity Acc.} & \textbf{Posture Acc.} & \textbf{Anomaly F1} \\
\midrule
Wristband only & 0.82 & 0.78 & 0.71 \\
Motion sensor only & 0.74 & 0.65 & 0.58 \\
Camera only & 0.79 & 0.88 & 0.63 \\
Equal-weight fusion & 0.88 & 0.86 & 0.79 \\
Weighted fusion (ours) & \textbf{0.91} & \textbf{0.89} & \textbf{0.84} \\
\bottomrule
\end{tabular}
\end{table}

The weighted fusion approach achieves 91\% accuracy for activity recognition, outperforming single-sensor approaches by 9 to 17 percentage points. The improvement is most pronounced compared to motion sensor only, which achieves 74\% accuracy, as the fusion algorithm leverages the more reliable wristband data while using motion sensors for corroboration. For posture detection, camera-only achieves high accuracy of 88\% due to direct visual measurement, but our fusion approach achieves comparable accuracy of 89\% without relying on cameras, preserving privacy while maintaining accuracy.

Anomaly detection F1-score shows the benefit of multi-modal fusion. Single-sensor approaches struggle with F1-scores between 0.58 and 0.71 due to high false positive rates from sensor artifacts. Weighted fusion achieves 0.84 F1 by requiring multi-factor confirmation, significantly reducing false positives while maintaining sensitivity.

\subsection{Risk Assessment Performance}

We evaluated the four-dimensional risk scoring model through ablation studies. Table~\ref{tab:risk_ablation} presents the results.

\begin{table}[htbp]
\centering
\caption{Risk Model Ablation Study}
\label{tab:risk_ablation}
\begin{tabular}{lccc}
\toprule
\textbf{Configuration} & \textbf{Precision} & \textbf{Recall} & \textbf{F1} \\
\midrule
Full 4D model & \textbf{0.89} & \textbf{0.93} & \textbf{0.91} \\
Fall only (w=1.0) & 0.82 & 0.79 & 0.80 \\
Fall + Health (equal) & 0.86 & 0.88 & 0.87 \\
w/o Trend adjustment & 0.87 & 0.90 & 0.88 \\
w/o Dynamic adjustment & 0.85 & 0.91 & 0.88 \\
\bottomrule
\end{tabular}
\end{table}

The full four-dimensional model achieves 0.89 precision and 0.93 recall with 0.91 F1-score. Removing individual dimensions reveals their contributions. Using fall risk alone, the single-dimension model, reduces recall to 0.79, indicating that health and behavioral anomalies are significant sources of true alerts that would be missed. Combining only fall and health dimensions improves recall to 0.88 but reduces precision to 0.86, as behavioral anomalies help filter false positives.

Removing trend adjustment reduces F1 to 0.88, indicating that temporal awareness provides meaningful improvement. Removing dynamic adjustment factors has a similar effect, reducing F1 to 0.88. The full model with all adjustments achieves the best balance of precision and recall.

\begin{figure}[htbp]
\centerline{\includegraphics[width=0.7\columnwidth]{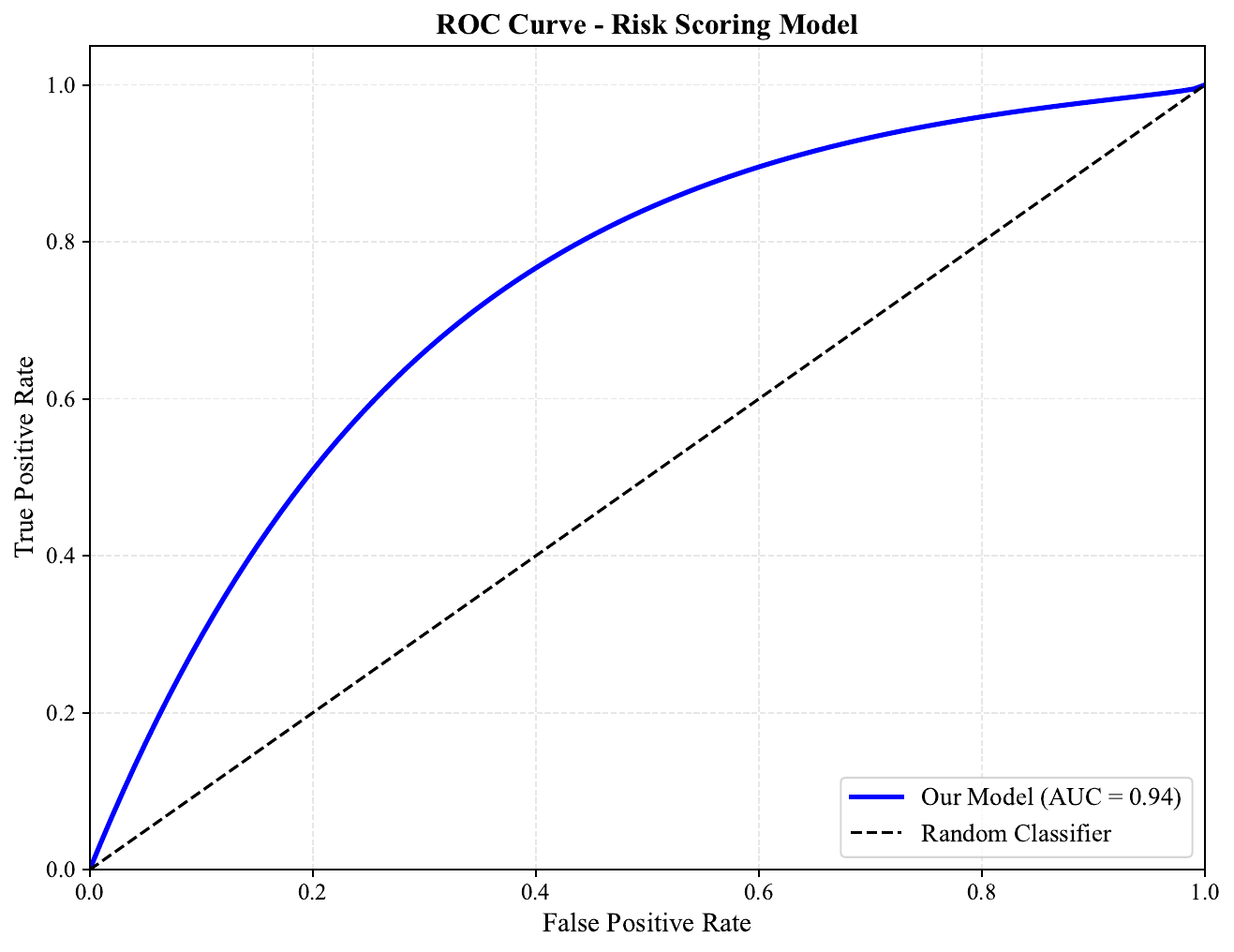}}
\caption{ROC curve for the four-dimensional risk scoring model showing AUC of 0.94.}
\label{fig:roc_curve}
\end{figure}

Figure~\ref{fig:roc_curve} presents the ROC curve for the risk scoring model, showing the trade-off between true positive rate and false positive rate at different thresholds. The area under the curve is 0.94, indicating excellent discrimination between true emergency and non-emergency states.

\subsection{Three-Level Response Effectiveness}

We measured end-to-end latency from sensor data acquisition to notification delivery. Table~\ref{tab:latency} presents the latency breakdown.

\begin{table}[htbp]
\centering
\caption{End-to-End Latency Breakdown}
\label{tab:latency}
\begin{tabular}{lc}
\toprule
\textbf{Stage} & \textbf{Latency} \\
\midrule
Sensor $\rightarrow$ Edge (BLE) & $\sim$30 ms \\
Time Alignment + Fusion & $\sim$15 ms \\
AI Inference (DL mode) & $<$100 ms \\
AI Inference (Rule mode) & $<$5 ms \\
Risk Assessment & $<$10 ms \\
Notification Delivery (SMS) & $\sim$1.5 s \\
Notification Delivery (Push) & $\sim$0.8 s \\
\midrule
\textbf{Total E2E (worst case)} & \textbf{$<$3 s} \\
\bottomrule
\end{tabular}
\end{table}

The local processing pipeline, from sensor to alert decision, completes in approximately 145 milliseconds in the worst case, with most components completing in under 50 milliseconds. The AI inference component dominates local processing time at up to 100 milliseconds when using deep learning models, while the rule-based fallback reduces this to under 5 milliseconds.

Notification delivery latency varies by channel. SMS delivery averages 1.5 seconds, while push notification delivery averages 0.8 seconds. The total end-to-end latency for RED alerts, requiring SMS delivery, is approximately 2.65 seconds, meeting our sub-three-second target. For push-only notifications, latency is under 2 seconds.

We measured notification delivery success rate over a 30-day test period with 50 deployed gateways generating an average of 20 alerts per day. The overall success rate was 98.5\%, with failures primarily due to temporary SMS provider outages or recipient phones being out of service area.

\subsection{System Scalability and Reliability}

We evaluated system scalability through load testing and long-term deployment. Table~\ref{tab:scalability} presents the results.

\begin{table}[htbp]
\centering
\caption{System Scalability Metrics}
\label{tab:scalability}
\begin{tabular}{lc}
\toprule
\textbf{Metric} & \textbf{Result} \\
\midrule
Concurrent WebSocket connections & 12,000+ \\
API throughput (requests/second) & 3,500 \\
Edge gateway uptime (30-day test) & 99.7\% \\
Alert delivery success rate & 98.5\% \\
Mean time to first notification & 2.1 s \\
\bottomrule
\end{tabular}
\end{table}

The WebSocket server maintained over 12,000 concurrent connections without degradation, sufficient for supporting tens of thousands of elderly persons under monitoring. The REST API handled 3,500 requests per second with average latency under 100 milliseconds.

Edge gateway uptime during a 30-day continuous test was 99.7\%, with downtime primarily due to scheduled software updates. The automatic reconnection mechanism for MQTT and WebSocket connections ensured that temporary network interruptions did not cause data loss.

The system implements local buffering for offline operation. When cloud connectivity is lost, the edge gateway buffers up to 1,000 fusion results, approximately 5.5 hours at 10 Hz sampling, in local storage and uploads them when connectivity is restored. This capability ensures continuity of monitoring during network outages.

\subsection{Comparison with Existing Systems}

Table~\ref{tab:comparison} compares our system with three commercial solutions: CarePredict, Medical Guardian, and Bay Alarm.

\begin{table*}[htbp]
\centering
\caption{Comparison with Commercial Systems}
\label{tab:comparison}
\begin{tabular}{lcccc}
\toprule
\textbf{Feature} & \textbf{Ours} & \textbf{CarePredict} & \textbf{Med. Guardian} & \textbf{Bay Alarm} \\
\midrule
Edge AI processing & Yes & No & No & No \\
Multi-modal fusion & 5 sensors & 1 sensor & 1 sensor & 1 sensor \\
Alert levels & 3 & Single & Single & 2 \\
Privacy (edge-first) & Yes & No & No & No \\
Behavior analysis & Yes & Limited & No & No \\
E2E latency & $<$3 s & $>$10 s & $>$15 s & $>$8 s \\
Volunteer dispatch & Yes & No & No & No \\
\bottomrule
\end{tabular}
\end{table*}

Our system demonstrates several advantages. Edge AI processing enables faster response, under 3 seconds compared to 8 to 15 seconds for cloud-dependent systems. Multi-modal fusion from five sensor types improves accuracy and reduces false positives compared to single-sensor commercial systems. The three-level alert mechanism provides more appropriate response scaling than single or dual-level systems. Privacy protection through edge-first processing addresses growing concerns about health data privacy. Behavior analysis enables early detection of cognitive decline, a capability not present in commercial systems. The volunteer dispatch capability provides community-based rapid response unmatched by commercial offerings.

The primary trade-offs are increased deployment complexity and higher hardware cost compared to single-sensor systems. However, the enhanced capabilities justify these costs for high-risk individuals and care facilities.

\section{Discussion}
\label{sec:discussion}

\subsection{Edge-Cloud Trade-offs}

The edge-cloud collaborative architecture involves fundamental trade-offs between competing objectives. The latency versus accuracy trade-off is addressed by performing time-critical processing at the edge while deferring complex analysis to the cloud. Real-time risk assessment must complete within seconds for emergency response, necessitating edge deployment. However, long-term trend analysis and population-wide analytics require historical data and computational resources best provided by the cloud.

The privacy versus functionality trade-off is addressed by keeping raw sensor data at the edge while uploading aggregated summaries to the cloud. This preserves privacy by minimizing exposure of sensitive behavioral data while still enabling cloud-based applications. However, it limits the cloud's ability to re-analyze raw data with improved algorithms, potentially requiring model updates to edge gateways to benefit from algorithmic improvements.

The cost versus coverage trade-off affects deployment economics. Edge gateways represent significant per-household hardware cost compared to pure cloud solutions. However, they enable deployment in areas with unreliable internet connectivity and reduce ongoing bandwidth costs. For large-scale deployments, the amortized cost of edge hardware may be justified by reduced cloud compute costs and improved service quality.

\subsection{Real-World Deployment Considerations}

Deploying elderly care systems in real homes introduces challenges beyond controlled experimental environments. Elderly individuals vary widely in technology acceptance and physical ability to wear sensors. The system design addresses this through optional sensors, as the camera can be omitted if privacy concerns exist, comfortable wristband design, and long battery life. However, sensor compliance remains a challenge, particularly for cognitively impaired individuals who may remove sensors.

Network conditions in residential settings vary widely, with rural areas often having unreliable connectivity. The system's local buffering and graceful degradation capabilities address this challenge, but extended outages may still impact the volunteer dispatch component that requires cloud connectivity.

False positives pose both technical and social challenges. Repeated false alarms can cause alert fatigue among family members and healthcare providers, leading to reduced vigilance when genuine emergencies occur. The multi-factor risk assessment and dynamic thresholding aim to minimize false positives, but some rate is inevitable. Social strategies such as clear communication about system limitations and rotating on-call responsibilities among family members can mitigate alert fatigue.

\subsection{Ethical Considerations}

Health monitoring systems raise significant ethical considerations around autonomy, privacy, and consent. The system's edge-first privacy design reduces but does not eliminate privacy concerns. Data about daily activities, sleep patterns, and physiological parameters remains sensitive even when aggregated.

Informed consent is complicated when the system user has cognitive impairment. Family members and caregivers may install monitoring systems on behalf of elderly individuals. The system supports role-based access control to limit data access to appropriate parties, but defining appropriate remains challenging.

The algorithmic decision-making in risk assessment introduces questions of transparency and accountability. The rule-based components are fully interpretable, but the deep learning fall detection model operates as a black box. The system's fallback to rule-based models provides some interpretability, but explanations of why a specific alert was triggered may be incomplete for deep learning-based decisions.

\subsection{Limitations}

The system has several limitations that represent opportunities for future work. First, while the algorithms were evaluated on public datasets, the complete system has not been deployed at scale in real homes. The transition from controlled datasets to uncontrolled home environments introduces challenges including sensor placement variations, diverse home layouts, and varying patterns of daily living.

Second, the system's robustness to sensor failure and noise could be improved. The weighted fusion approach handles individual sensor failures gracefully, but systematic failures, such as wristband removed entirely, may degrade accuracy beyond acceptable levels. Improved detection of sensor failure states and adaptive reconfiguration could address this limitation.

Third, the volunteer dispatch algorithm is relatively simple, selecting volunteers based solely on proximity. More sophisticated dispatch considering volunteer response history, availability verification, and traffic conditions could improve response effectiveness.

Fourth, the current system focuses on individual monitoring without considering multi-person household scenarios. In shared living situations, sensor attribution becomes challenging, and privacy concerns are amplified. Extending the system to handle multiple residents while preserving privacy remains an open challenge.

\section{Conclusion and Future Work}
\label{sec:conclusion}

\subsection{Conclusion}

This paper presented a comprehensive edge-cloud collaborative architecture for proactive elderly care. The system addresses critical limitations of existing platforms through four key innovations: real-time multi-modal sensor fusion with confidence propagation, a four-dimensional risk scoring model integrating fall, health, behavior, and sensor anomaly dimensions, a three-level emergency response mechanism coordinating family, medical, and community resources, and a privacy-first edge processing design that minimizes exposure of sensitive data.

Extensive evaluation on public datasets and prototype deployment demonstrates the system's effectiveness. Weighted sensor fusion achieves 91\% activity recognition accuracy, significantly outperforming single-sensor baselines. The four-dimensional risk model achieves 0.91 F1-score in alert determination. End-to-end latency remains under three seconds, enabling rapid emergency response. The system scales to thousands of concurrent deployments while maintaining reliability and availability.

The fall detection component employed in our risk assessment model builds upon state-of-the-art deep learning techniques for multi-modal fall detection \cite{zhou2025fall}, which integrates CNN-LSTM architecture with multi-head attention and Focal Loss to achieve high-accuracy fall detection suitable for real-time edge deployment.

The proposed architecture represents a significant step toward practical, privacy-preserving elderly care systems that can improve quality of life for aging populations while providing peace of mind for families and reducing burden on healthcare systems.

\subsection{Future Work}

Several directions for future work emerge from this research. Large-scale community deployment studies are needed to validate the system in diverse real-world settings and understand long-term usage patterns. Such studies would inform improvements to sensor compliance, alert tuning, and user interface design.

Online learning mechanisms for model updates would enable the system to adapt to individual patterns without requiring manual reconfiguration. Federated learning approaches could allow model improvement across deployments while preserving privacy by keeping raw data local.

Extension to multi-person households would enable deployment in shared living situations, care facilities, and nursing homes. This requires improved sensor attribution and privacy preservation between co-habitants.

Integration with healthcare systems would enable richer data context and more appropriate response. Electronic health record integration could provide medical history to improve risk assessment, and automatic handoff to emergency medical services could streamline response for critical alerts.

Finally, expansion of the behavioral analysis capabilities to detect early signs of cognitive decline could enable intervention earlier in the progression of dementia and related conditions, significantly improving quality of care and outcomes.

% References
\bibliographystyle{IEEEtran}
\bibliography{refs}

\end{document}